\documentstyle[12pt,epsf,epsfig,wrapfig2]{article}
\def\b1sum{\stackrel{\rule{.16in}{0.007in}}\sum}
\def\c{ \!\cdot\! }
\def\prod(#1,#2){ (#1\c #2) }

\def\U{ {\cal P} }

\def\W{ {\cal W} }

\def\X{ {\cal X} }

\def\tr{ {\rm Tr} }
\def\bk{\bar k}
\def\E{\bar E_k}
\def\md{ M_d }

\def\vk{\vec k}
\def\k{|\vk|}
 
\def\m{ \mu } 
\def\n{ \nu } 
 
\def\r{ \rho } 
\def\l{ \lambda } 
\def\rl{ {\rho\lambda} } 
\def\q{ \delta }

\def\g{ \gamma }
\def\thru#1{\mathrel{\mathop{#1\!\!\!/\!}}}
\def\s#1{\thru{#1}}

\def\equ(#1){Eq.\ (\ref{#1})}
\def\bea{\begin{eqnarray}}
\def\eea{\end{eqnarray}}
\def\be{\begin{equation}}

\def\ee{\end{equation}}
\def\z(#1){  \s{#1} + m }
\def\zm(#1){ \s{#1} - m }
\def\kp{{k}_+}
\def\km{{k}_-}
\def\fp{f_+}
\def\fm{f_-}
\def\us{ u(\vec k,s) }

\def\vs{ v(-\vec k,s) }

\def\a2{\alpha^2}

\begin{document}

\vspace*{-15mm}
\hspace*{\fill} ADP-96-38/T235\\

\begin{center}
{\large \bf
Relativistic effects in parity-violating electron-deuteron 
scattering and strangeness in the nucleon.
\\ }
\vspace{5mm}
Grigorios I.~Poulis$^{a,b}$
\\
\vspace{5mm}
{\small\it
(a) Department of Physics and Mathematical Physics\\ 
and Institute for Theoretical Physics\\ 
University of Adelaide, 5005 Adelaide SA, 
Australia~\footnote{present address}\\
and\\
(b) NIKHEF, P.O. Box 41882, 1009 DB Amsterdam, The Netherlands\\ }
\end{center}

\begin{center}
ABSTRACT

\vspace{5mm}
\begin{minipage}{130 mm}
\small
Deuteron electrodisintegration is described in a relativistic
PWIA framework. The reduction to the factorized PWIA,
and the effect thereof on the  observables accessible in
parity-violating scattering, is discussed. The extent to which
such relativistic effects may affect the extraction of
strangeness nucleon form factors from measurements of the 
PV asymmetry, is investigated.
\end{minipage}
\end{center}
Parity-violating (PV) inclusive scattering of 
polarized electrons off unpolarized protons
or nuclei provides a tool for extracting the hitherto unknown
strangeness nucleon form factors $G^s_{E,M}$.
The SAMPLE experiment at MIT-Bates~[1], a $Q^2$=$0.1$ (GeV/c)$^2$, 
backward angles measurement of the $p(\vec e,e')p$ asymmetry, is 
focusing on $G^s_M$. However, strong correlation with
the neutral current, axial, isovector form 
factor, $\tilde G_A^{T=1}$, whose radiative 
corrections have a large theoretical uncertainty in
the Standard Model, places constraints on the
effectiveness of the SAMPLE measurement. 
It has been suggested~[2] 
that a complimentary measurement of the 
PV asymmetry in quasielastic electron-deuteron scattering,
where this correlation is suppressed by a factor of 5,
may, in conjunction with the proton measurement, 
allow one to determine both $G^s_M$ and $\tilde G_A^{T=1}$.
A second phase of the SAMPLE experiment, as well as
a TJNAF (ex-CEBAF) experiment will perform such
measurements on the deuteron~[1]. 
Hadjimichael et al.~[2] have estimated the 
unavoidable theoretical uncertainty 
that nuclear dynamics introduces in the modeling of the
$^2H(\vec e,e')np$ reaction and
found that relativistic, rather than rescattering effects,
incorporated via factorized (F) PWIA,
are the main source of model dependence in the PV asymmetry 
for quasifree kinematics at $q>0.7$ GeV/c, 
especially at forward angles (6\% effects). 
In this work we treat the $^2H(\vec e,e')np$ reaction
in relativistic (R) PWIA, where not only the single-nucleon
tensor (as in FPWIA), but also the d-n-p vertex is treated relativistically.
To the extent that for quasifreee kinematics the 
two scales governing the relativistic effects
in this reaction, $q/m$ and $|\vk|/m$, decouple (Fig.~1), 
one expects that, since $|\vk|$ is restricted by 
the small ($\approx 60$ MeV/c) deuteron Fermi momentum, 
at high $q$ values the FPWIA employed in~[2]  
(although it does not treat relativistic effects 
associated with the $|\vk|/m$ scale properly) should not
differ significantly from the RPWIA. 

\begin{wrapfigure}{r}{5.5cm}
\centerline{
\epsfig{figure=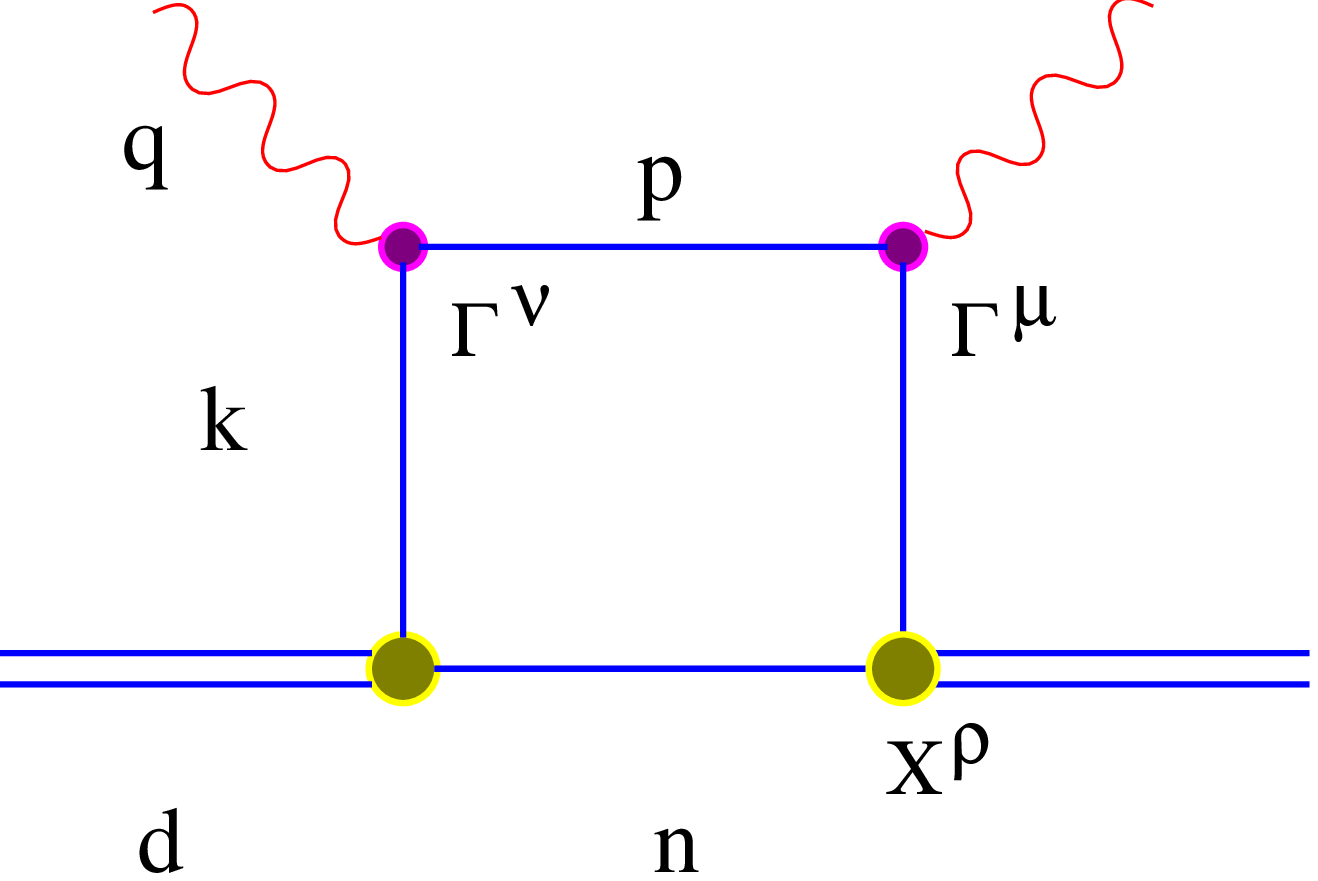,width=5.5cm,angle=-90}}
{\small Fig.~1.~PWIA hadronic tensor.}
\end{wrapfigure}

\noindent
 The purpose of this work is to test this conjecture. 
 The kinematics (deuteron rest frame) 
are described in Fig.~1; $q^\mu=(\omega,\vec q)$,
$d^\mu$ = $(M_d,\vec 0)$, $k^\mu$ = $(E_k,\vec k)$,
$n^\mu$ = $(E_n,-\vec k)$, $p^\mu$ = $q^\mu\!+\!
k^\mu$ = $(E_p,\vec p)$, with $d^2$ = $M_d^2$, 
$p^2$ = $n^2$ = $m^2$, $k^2$ $<$ $m^2$ 
(off-shell struck nucleon). In the (semirelativistic) 
FPWIA the hadronic tensor $W^{\mu\nu}_F$ 
factorizes as a nonrelativistic (NR) momentum distribution
$\rho(\k)$ times a relativistic off-shell 
single-nucleon tensor $\W^{\mu\nu}_F$:
\be\label{snt}
    W^{\m\n}_{F} = (2\pi)^3 M_d\;
 \rho(\k)\; \W^{\mu\nu}_F \ , \;\;
{\rm with}\;\;\W^{\mu\nu}_F =
 \tr\left[ (\z(\bk))\bar 
          \Gamma^\mu(\z(p))\Gamma^\nu\right](2m)^{-2} \ .
\ee
Here $\Gamma_\mu$ is the electromagnetic (E/M) 
or neutral current (NC) nucleon vertex, 
$\bar\Gamma_\mu$=$\gamma_0\Gamma_\mu^\dagger\gamma_0$,
and the de Forest~[3] prescription is used -- i.e., 
the $(\thru \bk +m)$ projector results from treating the struck nucleon
as ``would-be-on-shell'', using $\bk^\m$ = $(\E,\vk)$, 
where $\E^2$ = $E^2_n$ = $m^2\!+\!\vk^2$. In RPWIA, on the other hand, 
one uses the covariant d-n-p vertex $\X^\mu$ = 
$[ A\g^\mu + B \q^\mu $ $+$ $(m -\s k)
          (F\g^\mu + G \q^\mu)]$,
where $\q$=$(n\!-\!k)/2$, and the vertex functions $A,B,F,G$
can be written in terms of the  $^3S_1$, ($u$)
$^3D_1$, ($w$), $^3P_1$, ($v_t$), $^1P_1$ ($v_s$) 
relativistic deuteron wavefunctions. In this work
we use the  wavefunctions of Ref.~[4], obtained 
from the spectator equation with  pseudoscalar $\pi NN$ coupling, 
which enhances the role of the $P$ states. The hadronic 
tensor reads in PWIA 
\be\label{proton3}
W^{\mu\nu}_{R} = {1\over 3(k^2\!-\!m^2)^{2}}
{\rm Tr}\left[{\s \U\! +\beta \over 2m}(\s k\!+m)
\bar \Gamma^\m{\s p + m\over 2m}\Gamma^\n(\s k\!+m) \right] \ ,
\ee 
where $(\s \U \!+\beta)$=$\left(-g_\rl\! +\! d_\r d_\l/\md^2\right) 
\X^\l (\s n\!- m)  \bar \X^\r $. 
In order to reduce the above, {\it non}-factorized RPWIA
to the FPWIA of \equ(snt), we split the struck nucleon 
propagator into $+$ and $-$ energy contributions:
$	\z(k) = \fp(\s{k}_+ + m) + \fm (\s{k}_-m))$, where 
(in the target rest frame)
$f_\pm$=$(E_k \pm \E)/2\E$, $\kp=\bk$, $\km=n$. We then find
that a {\it weak}, density-matrix type of factorization still holds
(details will appear elsewhere) 
\be\label{density} 
	W^{\mu\nu}_{R} = (2\pi)^3 M_d\;{\textstyle \sum_{spins}}\left[
       	\rho_{++}\W^{\m\n}_{++} + 
			\rho_{+-}\W^{\m\n}_{+-} + 
			\rho_{-+}\W^{\m\n}_{-+} + 
			\rho_{--}\W^{\m\n}_{--}\right]  \ ,
\ee
where, in the shorthand notation $\chi^s_+=\us$ and $\chi^s_-=\vs$,
the generalized, relativistic ``momentum distributions''
and ``single-nucleon'' tensors are defined as
$\rho^{s,r}_{ab}$ $\sim$ $f_a f_b (k^2\!-\!m^2)^{-2}$
	           $ \tr [ \chi^s_a \bar\chi^r_b 
		               (\thru\U + \beta)]$
and
$ W^{\m\n}_{ab}$ = $\tr[ \chi^r_b  \bar \chi^s_a 
			 \bar\Gamma^\m  (\z(p))  
			  \Gamma^\n]/2m$, respectively.
Evaluating the traces we find that the $++$ and $--$ spectral
functions are diagonal in spin space,
$\rho^{s,r}_{\pm\pm}\sim \delta_{s,r}f_\pm^2
(k^2\!-\!m^2)^{-2}
[\prod(k_\pm,\U)/m \pm \beta]$, as well,
which then allows one to identify 
$\delta_{r,s}\W^{\mu\nu}_{++}=\W^{\mu\nu}_F$, 
[cf. \equ(snt)]. Furthermore, by taking the 
NR limit of $\rho_{++}$ (which is oblivious 
to the ``hard'' scale $q/m$), i.e., by ignoring 
the $P$ states and expanding in $|\vk|/m$, we find 
$\rho_{++}$ $\rightarrow$ $[u^2(|\vk|)+w^2(|\vk|)]/4\pi$ 
$=$ $\rho(|\vk|)$, the nonrelativistic momentum distribution
in \equ(snt). Thus, the widely 
used de Forest FPWIA may be obtained from 
RPWIA in a two-step reduction: (a) truncation to 
the $++$ components of the struck nucleon, 
(b) additionally, taking the nonrelativistic limit of 
the $\rho_{++}$ spectral function.

\begin{wrapfigure}{l}{8.3cm}
\epsfig{figure=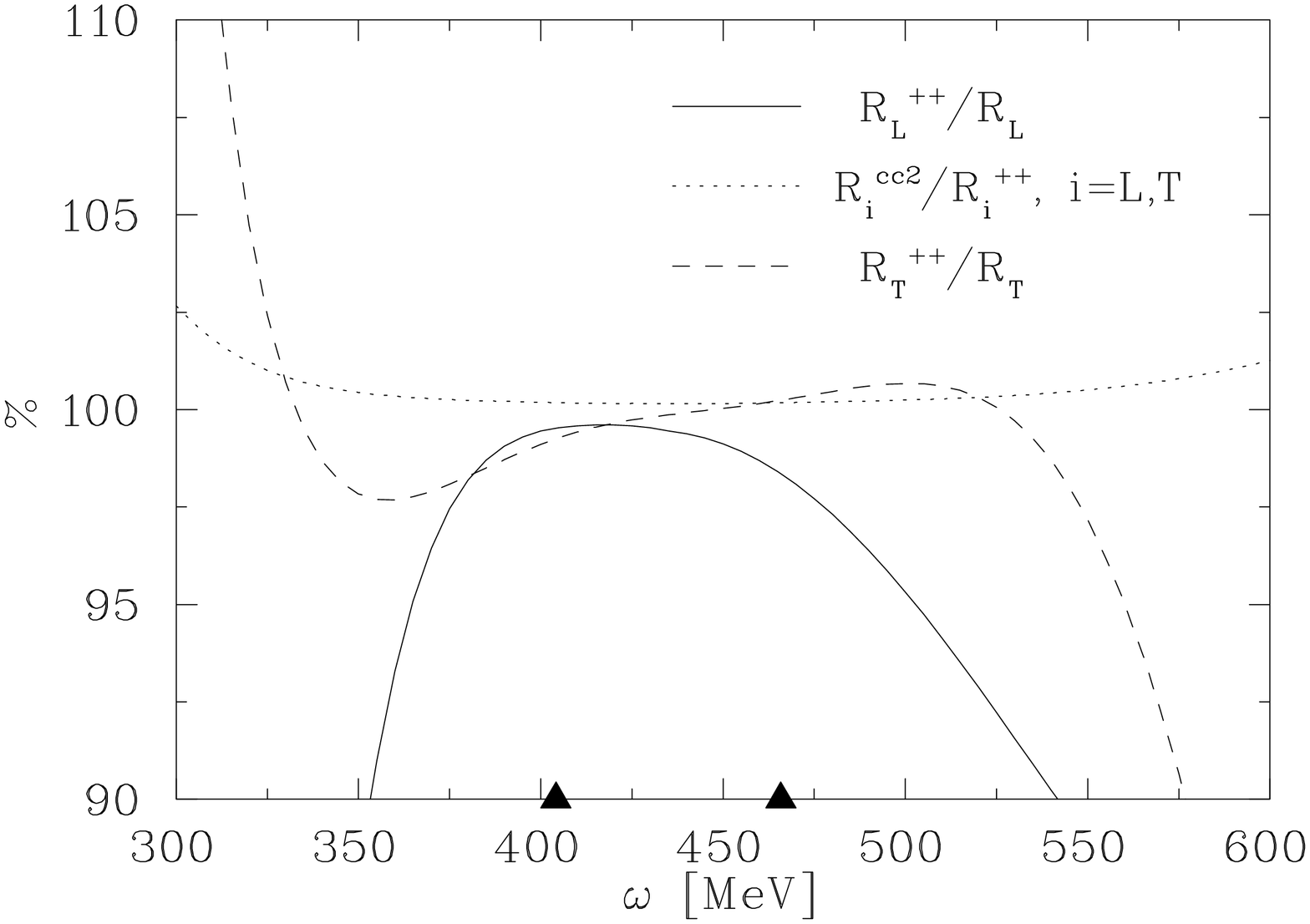,width=8.3cm,angle=90}
{\small Fig. 2. E/M responses at $q=1$ GeV/c.} 
\end{wrapfigure}

\noindent
The effects of these steps are examined
 in Fig.~2 for the longitudinal and transverse E/M 
responses at a fairly large value of momentum transfer, 
with enforced current conservation (``cc'')
and using the Dirac/Pauli (``2'') form
of the $\Gamma^\mu$ vertex. 
For a wide range of energy momentum values $\omega$,
step (b) induces a  $\le2\%$ effect, as seen from the 
$R^{cc2}_{L,T}/R^{++}_{L,T}$ ratio. Step (a), however, can be 
more severe, though not for quasifree kinematics: 
inside the quasielastic ridge (marked by the solid triangles) 
the ratio $R^{++}_{L,T}/R_{L,T}$ is within 1-2\%. 
The results show that relativistic effects
in inclusive, quasielastic scattering
are incorporated in the FPWIA, which should indeed be used at
large $q$ values, where a comparison with NR
treatments shows sizable effects,
as mentioned above. Relativistic effects associated with 
negative energy components of the off-shell struck nucleon
and with the target wavefunction are (in the target rest frame!) 
negligible for such kinematics. 
\begin{wrapfigure}{r}{8.0cm}
\epsfig{figure=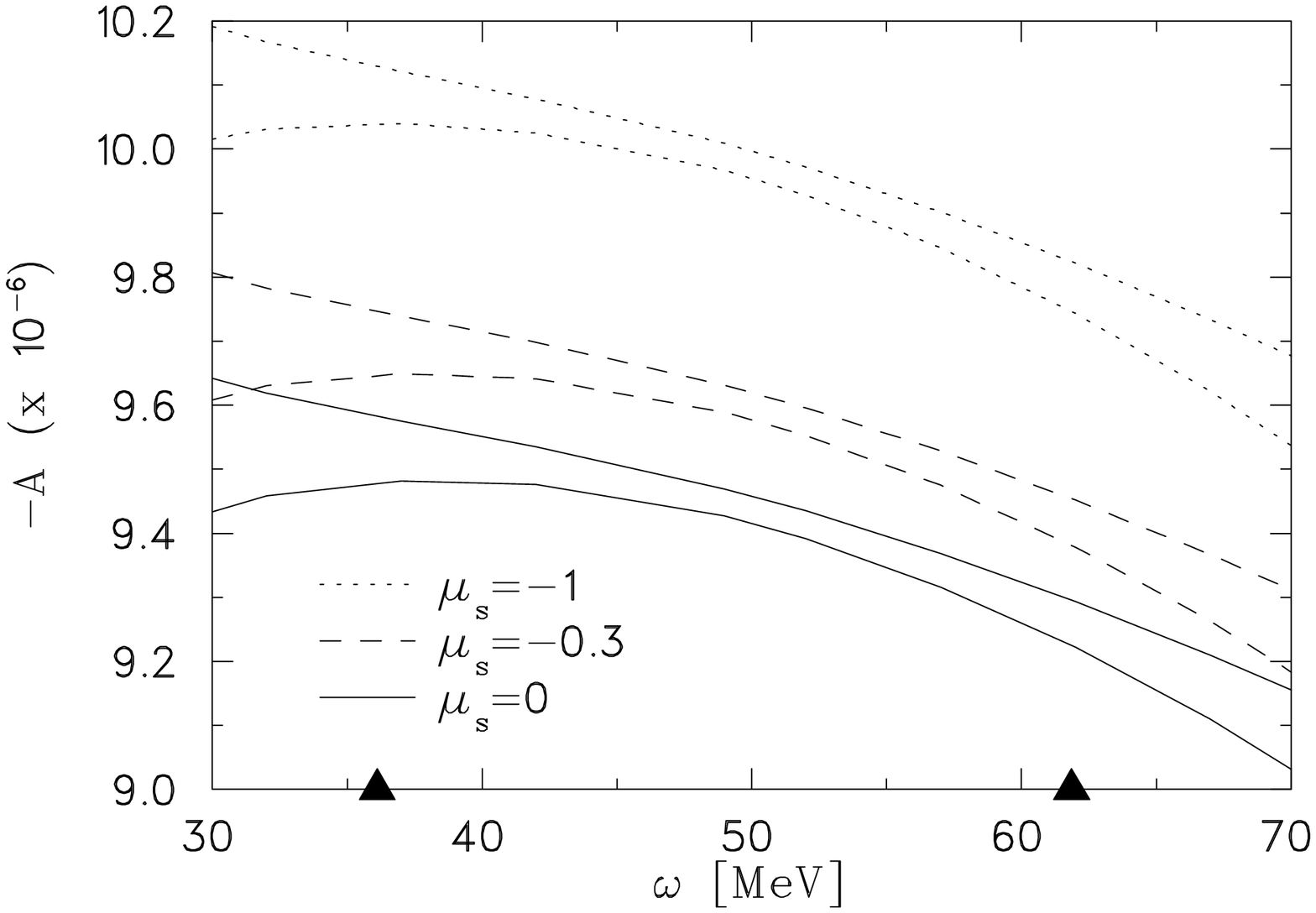,width=8.0cm,angle=90}
{\small  Fig.~3. Asymmetry at $q=300$ MeV/c. 
       Lower curves: RPWIA. Upper curves: FPWIA}.
\end{wrapfigure}

\noindent
Since we are primarily
interested in the interplay between the relativistic effects
{\it not} incorporated in the FPWIA and the extraction of 
strangeness, we show in Fig.~3 the PV asymmetry for a 
SAMPLE-type experiment, with three values of 
strangeness magnetic moment $\mu_s$~[1]. Interestingly enough,
it turns out that relativistic effects  beyond FPWIA in 
the quasielastic asymmetry are maximal
for $200 <q< 300$ MeV/c and backward angles, 
e.g, at the SAMPLE kinematics, where they induce a 0.45\% effect,
while 
$A=A_0(1-0.055\mu_s)$.
The extraction of strangeness on the quasielastic peak is 
nevertheless not obscured by such residual effects unless $\mu_s$ is quite 
small ($|\mu_s|<0.1$).
 
{\small\begin{description}

\item{[1]} E.J. Beise et al., archive: nucl-ex/9602001;  
                 also E.J. Beise, these proceedings.

\item{[2]} E. Hadjimichael, G.I. Poulis and T.W. Donnely, 
             Phys. Rev. C45, 2666 (1992).

\item{[3]} T. de  Forest, Nucl. Phys. A392, 232 (1983).

\item{[4]} W.W. Buck and F. Gross, Phys. Rev. C20, 2361, 
                (1979).


\end{description}}

\end{document}